\documentclass[apjl]{emulateapj} 
\usepackage{apjfonts}

\newcommand{\nuc}[2]{${}^{#2} \rm #1$}

\newcommand{\Eej}{\langle E_{\rm ej} \rangle}

\newcommand{\mev}{\, {\rm MeV} }

\newcommand{\Ms}{M_{\odot}}

\slugcomment{Submitted to Astrophys. J. Letters} 
\shorttitle{Li PRODUCTION ON A LOW-MASS SECONDARY IN BHSXT}
\shortauthors{FUJIMOTO, MATSUBA \& ARAI}

\begin{document}

\title{Lithium production on a low-mass secondary in a black hole soft X-ray transient}

\author{
Shin-ichiro Fujimoto\altaffilmark{1},
Ryuichi Matsuba\altaffilmark{2}
and Kenzo Arai\altaffilmark{3}
}

\altaffiltext{1}{
Department of Electronic Control, 
Kumamoto National College of Technology, 
2659-2 Suya, Koshi, 
Kumamoto 861-1102, Japan;
fujimoto@ec.knct.ac.jp}

\altaffiltext{2}{
Institute for e-Learning Development, 
Kumamoto University, Kumamoto 860-8555, Japan}

\altaffiltext{3}{
Department of Physics, 
Kumamoto University, Kumamoto 860-8555, Japan}


\begin{abstract}
We examine production of Li on the surface of a low-mass secondary in a black hole soft X-ray 
transient (BHSXT)
through the spallation of CNO nuclei 
by neutrons which are ejected from a hot (> 10 MeV) advection-dominated accretion flow (ADAF) around 
the black hole.
Using updated binary parameters, cross sections of neutron-induced spallation reactions,
and mass accretion rates in ADAF derived from the spectrum fitting of 
multi-wavelength observations of quiescent BHSXTs,
we obtain the equilibrium abundances of Li by equating the production rate of Li
and the mass transfer rate through accretion to the black hole. 
The resulting abundances are found to be 
in good agreement with the observed values in seven BHSXTs.
We note that the abundances vary in a timescale longer than a few months in our model.
Moreover, the isotopic ratio \nuc{Li}{6}/\nuc{Li}{7} 
is calculated to be about 0.7--0.8 on the secondaries, which is much higher than the ratio measured 
in meteorites.
Detection of such a high value is favorable to the production of Li via spallation
and the existence of a hot accretion flow, rather than an accretion disk corona system in quiescent BHSXT.
\end{abstract}


\section{Introduction}


High abundances of Li have been detected 
in late-type secondaries of black hole soft X-ray transients 
(BHSXTs) and a neutron star soft X-ray transient (NSSXT) in quiescence~\citep{martin92,martin94,martin96},
though Li would be destructed in a deep convective envelope of a late-type star. 
The Li enrichment has not, however, been observed on a late-type secondary 
in a compact binary with a white dwarf~\citep{martin95}.
These facts strongly suggest that a production mechanism of Li
operates in compact binaries~\citep{yn97,gk99} and that the nature of the primaries
is crucial for the mechanism,
though rotation might reduce the destruction of Li in the envelope 
of the secondary~\citep{mjs05}.

Multi-wavelength spectra of BHSXTs in quiescence are 
successfully fitted to the radiation from an advection-dominated accretion flow (ADAF)
around the black hole~\citep{nmy96,nbm97}.
Density is so low in ADAF, that ions interact inefficiently with electrons.
Consequently ions have high temperatures due to viscous heating 
up to about 30 MeV near the inner edge of ADAF.
At such high temperatures,  $\alpha$-$\alpha$ reaction 
proceeds to synthesize Li inside ADAF~\citep{martin94,yn97}.
It is necessary that a fraction $10^{-3}-10^{-4}$ of the accreting gas
is transported to the secondary to explain the high abundances of Li
observed in BHSXTs.
However, such a high fraction is uncertain to be realized due to strong gravity 
of the black hole and the Coulomb interactions with nuclei inside ADAF~\citep{gk99}.

Helium breaks via spallation with protons to produce neutrons at the inner region of ADAF.
A large fraction of neutrons can be ejected from ADAF, because they 
do not interact with nuclei through the Coulomb interactions.
Neutrons intercepted by the secondary interact with CNO nuclei through spallation 
to produce Li on the surface~\citep{gk99}.
This scenario is of particular interest, 
because the Li enrichment is anticipated in secondaries only for BHSXTs and NSSXTs,
but for white dwarfs as primaries where ADAF cannot attain 
enough high temperatures to break helium to nucleons there.

In the present paper, we evaluate the Li abundances on the surface of secondaries in BHSXTs, 
following the scenario proposed by \citet{gk99}.
To this end, we use updated binary parameters, such as the mass $M$ of a black hole, 
the mass $M_*$ and radius $R_*$ of a secondary, 
mass accretion rates derived from the spectrum fitting of multi-wavelength observation of BHSXTs in quiescence,
and cross sections of neutron-induced spallation reactions.
Then, we compare the resulting abundances with the observed values, 
and show that the agreement is quite well.
Moreover, we predict the isotopic ratio \nuc{Li}{6}/\nuc{Li}{7} on the secondaries in BHSXTs.

\section{Neutron production in an advection dominated accretion flow}
\label{sec:nuc-adaf}

Temperature of ions in ADAF is comparable to virial temperature, and is given at radius $r$ 
by~\citep{ny94, ny95a, ny95b};
\begin{equation}
 T = 3.7 \times 10^{12} \frac{r_{\rm in}}{r} \, {\rm K} 
 = 31.9 \frac{r_{\rm in}}{r}\, {\rm MeV}. \label{eq:temp}
\end{equation}
Here $r_{\rm in}$ is the radius at the inner edge of ADAF, 
and is set to be $3 r_g$, 
where $r_g$ is the Schwarzschild radius of the black hole.
The number density is also given by 
\begin{equation}
 n = 1.7 \times 10^{18} \alpha^{-1} m^{-1} \dot{m} 
\left(\frac{r}{r_{\rm in}}\right)^{-3/2} \,{\rm cm^{-3}}, 
\label{eq:ndens}
\end{equation}
where $\alpha$ is the viscous parameter, $m = M/\Ms$, and $\dot{m}$ is
the mass accretion rate in units of the Eddington accretion rate
$\dot{M}_{\rm Edd} = 1.4 \times 10^{17} \, m \ \rm g\, s^{-1}$.

\begin{figure}
\epsscale{0.9}
\plotone{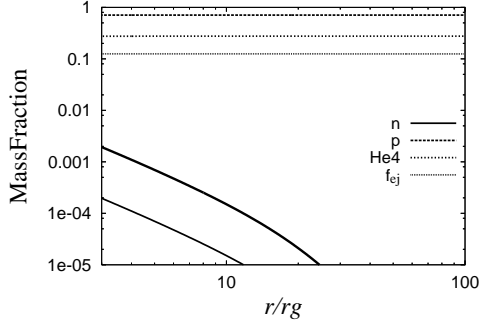}
\caption{Distribution of abundance in ADAF for
$\alpha = 0.3$, $m = 10$, and $\dot{m} = 10^{-3}$.
The solid, dashed, and dotted lines indicate 
the mass fractions of n, p, and \nuc{He}{4}, respectively.
The thick solid line denotes the neutron fraction for $\dot{m} = 10^{-2}$.
The ejection fraction $f_{\rm ej}$ of neutrons is also presented by 
the dot-dashed line.
} \label{fig:abund-adaf}
\end{figure} 

Once the temperatures, densities and drift timescales are specified, 
we can follow the abundance evolution in ADAF 
from the outer boundary $r_{\rm out}$ to $r_{\rm in}$, using a nuclear reaction network.
We set $r_{\rm out}$ to be $100 \, r_g$.
It is likely that $r_{\rm out}$ becomes much larger during the quiescent state~\citep{nbm97},
but the abundance of neutrons is independent from the choice of larger $r_{\rm out}$ 
because of low temperatures ($< 1\,\rm MeV$) in the outer region~\citep{gk99}.
Our network contains 17 species of nuclei; 
n, p, D, T, \nuc{He}{3}, \nuc{He}{4}, \nuc{B}{9}, \nuc{C}{11}, \nuc{C}{12},
\nuc{N}{13}, \nuc{N}{14}, \nuc{O}{15}, \nuc{O}{16}, \nuc{F}{17}, \nuc{Ne}{20}, 
\nuc{Na}{21},  and \nuc{Mg}{24},  
and 14 reactions, whose rates are taken from Table 1 in \citet{gg89}.
It should be emphasized that photodisintegration reactions are not important for 
abundance evolution inside ADAF, since ADAF is optically thin and photons have no 
chance to interact with nuclei due to low gas densities (Eq. (\ref{eq:ndens})).
Therefore, nuclear statistical equilibrium cannot be realized in ADAF even 
for high temperatures (Eq. (\ref{eq:temp})).
The network is appropriate for the study of the production of neutrons in ADAF, 
but insufficient for heavy nuclei as well as Li because of the limited numbers of nuclei and 
reactions.
Initial abundance at $r_{\rm out}$ 
is set to be the solar composition~\citep{ag89}.

Figure \ref{fig:abund-adaf} shows the abundance distribution inside ADAF
for $\alpha = 0.3$, $m = 10$, and $\dot{m} = 10^{-3}$.
Neutrons are produced significantly via the breakup of \nuc{He}{4} at $r < 20 r_g$.
The distribution of neutrons is similar to that in Figure 1 of \citet{jg01}.
We note that the number fraction of neutrons $Y_{\rm n}$ depends not on $m$ solely, but
on the combination $\dot{m}/\alpha^2$.
Hereafter we fix $\alpha = 0.3$ in the present paper~\citep{nbm97}.
It should be emphasized that the breakup of helium cannot take place in an accretion corona, 
which is an alternative scenario to explain multi wavelength spectrum of BHSXTs 
in quiescence~\citep[e.g.][]{malzac07}
because of low ion temperatures comparable to electron temperature ($< 1\,\rm MeV$).

The neutrons produced in ADAF have positive Bernoulli numbers~\citep{ny94}, so that
a fraction of the neutrons thermally overcomes the deep gravitational well of the black hole 
before inelastic scattering with protons.
The ejection fraction of neutrons $f_{\rm ej}$ is evaluated from Eq. (14) 
in \citet{gk90}, using the pseudo-Newtonian potential and
an experimentally measured cross section 
of the neutron-proton inelastic scattering, 
$\sigma_{\rm np} = 671.0 \, (14.1{\,\rm MeV}/E_{\rm n})$ mb \citep{tks70},
where $E_{\rm n}$ is the energy of neutrons. 
The distribution function of neutrons is set to be Maxwellian
with ion temperature of ADAF~\citep{gk90,gk99}.
We find that 
$f_{\rm ej} \simeq 0.12$, which depends weakly on $r$ as seen from Figure \ref{fig:abund-adaf}. 
It is noted that $f_{\rm ej}$ is independent of $\alpha$, $m$, and $\dot{m}$.

Using the mass conservation in ADAF, 
we evaluate the ejection rate of neutrons from ADAF as
\begin{eqnarray}
 \dot{N}_{\rm n}
 &=& \int^{r_{\rm out}}_{r_{\rm in}} f_{\rm ej} \left(\frac{dY_{\rm n}}{dt}\right) 
 \frac{2\pi r \Sigma }{m_n} dr \
 \simeq \frac{\dot{M}}{m_n} Y_{\rm n,in} f_{\rm ej,in} \nonumber \\
 &\simeq& 1.1 \times 10^{34}
  \left(\frac{\dot{m}}{10^{-3}}\right) 
  \left(\frac{m}{10}\right) 
  \left(\frac{f_{\rm ej,in}}{0.1}\right)
  \left(\frac{Y_{\rm n, in}}{10^{-4}}\right) \, {\rm s^{-1}},
\end{eqnarray}
where $\Sigma$ is the surface density in ADAF,
$f_{\rm ej,in}$ and $Y_{\rm n, in}$ are the values of $f_{\rm ej}$ and $Y_{\rm n}$ at $r_{\rm in}$.
Here we have used a relation $df_{\rm ej}/dr \simeq 0$ (see Figure \ref{fig:abund-adaf}).
We note that the rate is unlikely to change significantly for smaller $r_{\rm in}$.
Even if we set $r_{\rm in} < 3 r_g$, 
the increase in $Y_{\rm n,in}$ due to higher temperatures would be 
canceled out by a large decrease in $f_{\rm ej,in}$ resulted from general relativistic effects. 

Next we calculate the energy of the ejected neutrons
averaged over the region from $r_{\rm in}$ to $r_{\rm out}$ as
\begin{equation}
\Eej = 
\frac{1}{m_n \dot{N}_{\rm n}}
\int^{r_{\rm out}}_{r_{\rm in}} E_{\rm ej} f_{\rm ej} \left(\frac{dY_{\rm n}}{dt}\right) 2\pi r \Sigma dr, 
\end{equation}
where $E_{\rm ej}$ is the energy of neutrons ejected from ADAF, 
and is evaluated from the same way as in $f_{\rm ej}$.
It is noted that $\Eej$ is crucial for the production of Li on the secondary, 
because cross sections of both the spallation reactions and the inelastic scattering with protons
depend strongly on the neutron energy.
We find that $\Eej \simeq 78\mev$, which is insensitive to $\alpha$, $m$, and $\dot{m}$.

\section{Li production on the secondary through spallation of 
CNO nuclei by neutrons}\label{sec:Li-production}

The surface of a secondary in BHSXT is bombarded by neutrons from ADAF.
We note that $\beta$-decays of neutrons can be ignored,
because their half-life is much longer than 
the elapsed time $230\, (a/R_\odot)(0.1/v_{\rm ej}) \, \rm s$
during the flight from ADAF to the surface, where $a$ and $v_{\rm ej}$ are 
the binary separation and the ejection velocity in units of the velocity of light.
The depth of an envelope exposed by neutrons
is expressed as $1/n_{\rm p} \sigma_{\rm np}$, where $n_{\rm p}$ is
the number density of protons on the surface of the secondary,
since the neutron-proton inelastic scattering is predominant.
The mass of the neutron-exposed envelope is given by
\begin{eqnarray}
 M_{\rm exp} 
             &\simeq& 2.3 \times 10^{-10}
	      \left(\frac{R_*}{0.7 R_\odot} \right)^2 
	      \left(\frac{0.9}{Y_{\rm p}} \right)
	      \left(\frac{121\rm mb}{\sigma_{\rm np}} \right) \,
	      M_{\odot}. 
\end{eqnarray}
The abundance of Li increases through the spallation of CNO nuclei
by neutrons on the surface of the secondary.
For isotropic ejection of neutrons from ADAF, the production rate of Li 
on the secondary is given by
\begin{eqnarray}
 \dot{M}_{\rm Li}^{+} 
  &=& \frac{1}{2}\frac{\dot{N}_{\rm n}}{4\pi a^2} \sigma_{\rm sp} M_{\rm exp}
  Y_{\rm CNO} \bar{A}_{\rm Li} \nonumber \\
  &\simeq& 1.6 \times 10^{-20} 
  \left(\frac{\dot{m}}{10^{-3}}\right) 
  \left(\frac{m}{10}\right) 
  \left(\frac{f_{\rm ej,in}}{0.1}\right)
  \left(\frac{0.9}{Y_{\rm p}}\right) 
  \left(\frac{Y_{\rm n,in}}{10^{-4}}\right) \nonumber \\ 
  & & \hspace{-5mm} \times 
  \left(\frac{Y_{\rm CNO}}{10^{-3}}\right) 
  \left(\frac{\sigma_{\rm sp}}{25 \rm mb}\right)
  \left(\frac{121 \rm mb}{\sigma_{\rm np}}\right)
  \left(\frac{R_*}{0.25 a}\right)^2 
  \, M_{\odot} \, {\rm yr^{-1}}, 
  \label{eq:M_Li+}
\end{eqnarray}
where $\bar{A}_{\rm Li}$ is the average mass number of Li, 
which is composed of \nuc{Li}{6} and \nuc{Li}{7}, and is set to be 7, 
$\sigma_{\rm sp}$ is the total cross section of the spallation reactions of CNO nuclei,
and $Y_{\rm CNO}$ is the number fraction of CNO nuclei,
which is $1.2 \times 10^{-3}$ for the solar abundances~\citep{ag89}.
A factor 1/2 in Eq.~(\ref{eq:M_Li+}) means the fact that 
a half of the surface of the secondary is exposed by neutrons from ADAF.

On the other hand, a fraction of the produced Li is transported to the black hole through accretion.
The mass transfer rate of Li from the envelope is expressed as 
\begin{eqnarray}
 \dot{M}_{\rm Li}^{-} 
  &\simeq& 1.5 \times 10^{-20} 
   \left(\frac{Y_{\rm Li}}{10^{-10}}\right)
   \left(\frac{m}{10}\right) 
   \left(\frac{\dot{m}}{10^{-3}}\right) 
\, M_{\odot} \, {\rm yr^{-1}}.
\end{eqnarray}
It should be noted that the destruction rate of Li 
in the envelope is
$7 M_{\rm exp} Y_{\rm Li}/t_{\rm des}$, which 
is much smaller than $\dot{M}_{\rm Li}^{-}$ even for a short destruction timescale 
$t_{\rm des} \simeq 10^7 \rm yr$. 

For equilibrium between the production and loss rates $\dot{M}_{\rm Li}^+ = \dot{M}_{\rm Li}^-$, one can obtain
\begin{eqnarray}
 Y_{\rm Li, \, eq} 
  &\simeq& 1.1 \times 10^{-10} 
  \left(\frac{f_{\rm ej,in}}{0.1}\right)
  \left(\frac{Y_{\rm n, in}}{10^{-4}}\right) 
  \left(\frac{0.9}{Y_{\rm p}}\right) 
  \left(\frac{Y_{\rm CNO}}{10^{-3}}\right) \nonumber \\
  & & \times 
  \left(\frac{\sigma_{\rm sp}}{25 \rm mb}\right)
  \left(\frac{121 \rm mb}{\sigma_{\rm np}}\right)
  \left(\frac{R_*}{0.25 a}\right)^2 .
\end{eqnarray}
It should be emphasized that $Y_{\rm Li, \, eq}$ depends on $\alpha$ and $\dot{m}$
as the combination of $\dot{m}/\alpha^2$ through $Y_{\rm n,in}$.

The timescale for the Li enhancement is given by
\begin{eqnarray}
 \tau_{\rm eq}
  &=& M_{\rm exp} \bar{A}_{\rm Li} Y_{\rm Li, eq} /\dot{M}_{\rm Li}^{+} \nonumber \\
  &\simeq& 10.2
  \left(\frac{f_{\rm ej,in}}{0.1}\right)^{-1}
  \left(\frac{Y_{\rm Li,eq}}{10^{-10}}\right) 
  \left(\frac{Y_{\rm n,in}}{10^{-4}}\right)^{-1}
  \left(\frac{Y_{\rm CNO}}{10^{-3}}\right)^{-1} \nonumber \\
  & & \times 
   \left(\frac{m}{10}\right)^{-1}
  \left(\frac{\dot{m}}{10^{-3}}\right)^{-1}
  \left(\frac{a}{3 R_\odot}\right)^2 
  \left(\frac{25\rm mb}{\sigma_{\rm sp}} \right)
  \, {\rm yr}.
\end{eqnarray}
Therefore, it takes a few years for lithium to achieve the equilibrium abundance of $10^{-10}$.
We note that the Li abundance varies in a timescale longer than a few months in our model, 
while this is not the case in the scenario proposed by \citet{yn97}.


Finally, we evaluate 
the total cross sections of spallation reactions of CNO nuclei induced by neutrons
as the sum of the cross sections of 
\nuc{C}{12}, \nuc{N}{14}, and \nuc{O}{16} weighted by their number fractions.
The cross sections for \nuc{C}{12}, \nuc{N}{14}, and \nuc{O}{16} 
are calculated from the Talys nuclear reaction code~\citep{khd05}.
The energy distribution of neutrons just before the spallation is 
assumed to be the same as that of neutrons ejected from ADAF, 
because neutrons are unlikely to lose their energies largely 
during the flight.
For the solar abundance and $E_{\rm n} = \Eej = 78\mev$, 
we find that the cross sections
are $\sigma_{\rm sp6} = 10.5 \,\rm mb$ and $\sigma_{\rm sp7} = 15.3 \,\rm mb$ 
for the production of \nuc{Li}{6} and \nuc{Li}{7}, respectively, 
yielding the sum $\sigma_{\rm sp} = \sigma_{\rm sp6} + \sigma_{\rm sp7} = 25.8 \,\rm mb$. 
If we adopt the abundance of CNO-processed material, in which all the original \nuc{C}{12} are
converted to \nuc{N}{14} through CNO cycle in the interior of the secondary, 
the total cross section decreases to be $\sigma_{\rm sp} = 18.0 \,\rm mb$.
It is noted that the cross section of the spallation $\sigma_{\rm sp}$ is smaller
than that of the inelastic scattering $\sigma_{\rm np}$
by a factor 5 at $E_{\rm n} = 78\mev$.

\section{Comparison with observations}
\label{sec:comparison}

\begin{deluxetable} {lcccccc}
\tablewidth{0pt}
\tablecaption{Parameters of BHSXTs}
\tablehead{
\colhead{object} &
\colhead{$M$} &
\colhead{$M_*$} & 
\colhead{$a$} & 
\colhead{$P_{\rm orb}$}  &
\colhead{$A(\rm Li)_{\rm obs}$}\\
&($M_\odot$)&($R_\odot$)&($R_\odot$)&(day)&
}
\startdata
XTE J1118+480 
                & 6.8 & 0.25 & 1.4 & 0.171 &  <1.86 \\ 
GRO 0422+32\tablenotemark{a} 
                & 9   & 0.39 & 2.3 & 0.212 &  <1.62 \\ 
A0620-003\tablenotemark{b} 
                & 9.7 & 0.65 & 3.6 & 0.323 &  2.31$\pm$ 0.21 \\ 
QZ Vul 
                & 8.5 & 0.5  & 2.6 & 0.345 &  2.20$\pm$ 0.50 \\ 
GU Mus 
                & 6   & 0.8  & 3.3 & 0.433 &  3.00$\pm$ 0.50   \\ 
Nova Oph77 
                & 4.9 & 0.7  & 2.9 & 0.521 &  <2.96 \\ 
V404 Cyg\tablenotemark{c} 
                & 12  &  0.7 & 3.8 & 6.47  &  2.70$\pm$ 0.20
\enddata
\tablenotetext{a}{\citet{bradley07}}.
\tablenotetext{b}{\citet{froning07}}.
\tablenotetext{c}{\citet{esin98} and referense threin}.
\tablecomments{
$M$, $M_*$, and $P_{\rm orb}$ 
are taken from \citet{charles06} and \citet{chen97}.
$A(\rm Li)_{\rm obs} = \log(Y_{\rm Li}/Y_{\rm p}) + 12$ is adopted from table 5 in \citet{casares07}.
}
\end{deluxetable}

We compare the evaluated and observed abundances of Li in seven BHSXTs,
using updated parameters of the binaries in Table 1.
The radii of the secondaries are taken from the simulations
of the evolution of (single) spherical stars with corresponding masses
and the solar metallicity at 1Gyr~\citep[Table 2 in][]{cb97}; 
They vary from $0.21 R_\odot$ to $0.67 R_\odot$ 
as the secondary masses increase.
The binary separations are calculated from $R_*/a = 0.46 ( 1 + M/M_*)^{-1/3}$~\citep{p71}.
The observed abundances of Li are adopted from \citet{casares07}.
We note that an upper limit of $A(\rm Li)_{\rm obs}$ for GRO 0422+32 was evaluated 
as a higher value~\citep[2.0;][]{martin96}, instead of 1.62~\citep{casares07}, 
which has been obtained from the reanalysis of observational data by \citet{martin96}.

The mass accretion rates in ADAF are found from the spectrum fitting of 
the multi-wavelength observations of quiescent BHSXTs 
to be $\dot{m}_{\rm spe} = 4.3 \times 10^{-3}$ and $2.0 \times 10^{-2}$
for A0620-003 and V404 Cyg, respectively~\citep{nbm97,qn99}.
For the other objects, where the spectrum fitting has not yet been performed in quiescence,
we simply specify the accretion rates from
the minimum X-ray luminosities in these systems~\citep{g01,m03}.
These values of $\dot{m}_{\rm spe}$ are given in Table 2.

\begin{figure}[ht]
\epsscale{0.9}
\plotone{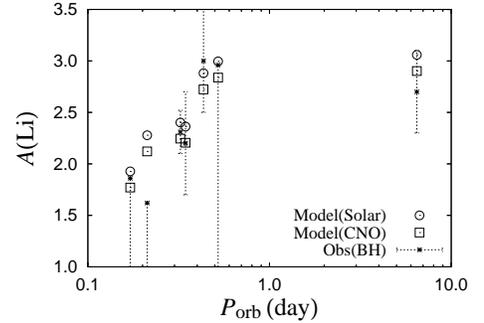}
\caption{Evaluated and observed abundances of Li on the secondaries in seven BHSXTs
with respect to their orbital periods.
} \label{fig:ALi}
\end{figure} 

Using these values of parameters all together, we calculate the equilibrium abundances of Li 
on the secondaries in seven BHSXTs.
The resulting abundances $A(\rm Li) = \log(Y_{\rm Li, eq}/Y_{\rm p}) + 12$ and  
the enhancement timescale $\tau_{\rm eq}$ are given in Table 2.
We show $A(\rm Li)$ against the orbital periods $P_{\rm orb}$  of the binaries by the open circles 
in Figure \ref{fig:ALi}.
It is found that our results are in good agreement with the observed abundances, 
in particular for $P_{\rm orb} \ge 0.3 \,\rm day$.

If we adopt the composition of CNO-processed material for the secondary,
we obtain lower abundances of Li, as denoted by the open squares in Figure \ref{fig:ALi}.
This is favorable to BHSXTs with $P_{\rm orb} < 0.3 \,\rm day$, 
such as XTE J1118+480~\citep{es01, haswell02}.

\begin{figure}[ht]
\epsscale{0.9}
\plotone{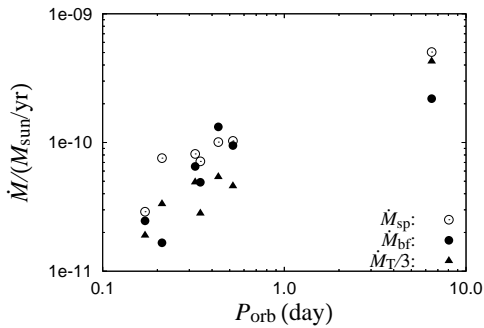}
\caption{Mass accretion rates with respect to the orbital periods.
} \label{fig:Mdot}
\end{figure} 

Next, we try to fit our results to the observed Li abundances with varying accretion rates.
The resulting best-fit rates $\dot{m}_{\rm bf}$ are given in Table 2.
It is found that they are comparable to or slightly lower than $\dot{m}_{\rm spe}$.
They are also nearly 1/3 of the rates $\dot{m}_T$ predicted by 
binary evolution models~\citep{king96}, 
where a factor 1/3 is taken into account with the accumulation of accreting material
in an outer thin disk~\citep{menou99}.
Figure \ref{fig:Mdot} shows the comparison of the mass accretion rates
$\dot{M}_{\rm spe}$, $\dot{M}_{\rm bf}$, and $\dot{M}_{\rm T}/3$
with respect to the orbital periods in BHSXTs.

Finally, we evaluate the isotopic ratio \nuc{Li}{6}/\nuc{Li}{7} on the secondaries.
It is easily calculated from the cross sections of \nuc{Li}{6} and \nuc{Li}{7} for the spallation reactions
to be 0.69 -- 0.81, depending on the CNO abundances of the secondaries.
We note that the ratio is much larger than 0.12 for NSSXT Cen X-4~\citep{casares07}
and 0.081 for meteorites~\citep{ag89}.
Detection of such a high \nuc{Li}{6}/\nuc{Li}{7} ratio will be an evidence for the production of 
Li on the secondaries in BHSXTs.

\begin{deluxetable} {lcccccc}
\tablewidth{0pc}
\tablecaption{Abundances of Li and accretion rates.}
\tablehead{
\colhead{object} &
\colhead{$A(\rm Li)$} & 
\colhead{$\dot{m}_{\rm spe}$} &
\colhead{$\dot{m}_{\rm bf}$} &
\colhead{$\dot{m}_{T}/3$} &
\colhead{$\tau_{\rm eq}$ (yr)}
}
\startdata
  XTE J1118+480 &  1.93 & 2.0e-3 & 1.7e-3 & 1.3e-3 &  0.63 \\ 
    GRO 0422+32 &  2.28 & 4.0e-3 & 8.8e-4 & 1.8e-3 &  0.44 \\ 
      A0620-003 &  2.40 & 4.3e-3 & 3.2e-3 & 2.4e-3 &  2.06 \\ 
         QZ Vul &  2.36 & 4.0e-3 & 2.7e-3 & 1.6e-3 &  0.94 \\ 
         GU Mus &  2.88 & 8.0e-3 & 1.1e-2 & 4.3e-3 &  3.40 \\ 
     Nova Oph77 &  3.00 & 1.0e-2 & 9.2e-3 & 4.5e-3 &  1.86 \\ 
       V404 Cyg &  3.06 & 2.0e-2 & 8.7e-3 & 1.7e-2 &  0.18
\enddata
\end{deluxetable}

\section{Summary} 
\label{sec:summary}

We have evaluated the Li abundances on the surface of the low-mass secondaries in quiescent BHSXTs,
using the updated parameters of the binaries and the cross sections of neutron-induced spallation reactions.
The mass accretion rates in ADAFs are derived from the spectrum fitting of multi-wavelength 
observations~\citep{nbm97,qn99} or specified from the minimum X-ray luminosities of BHSXTs. 
Significant amounts of Li are produced on the surface through the spallation of CNO nuclei 
by neutrons which are ejected from hot ADAFs around the black hole.
We have obtained the equilibrium abundances of Li by equating the production rate of Li
on the secondary 
and the mass transfer rate through accretion to the black hole. 
It is found that the resulting abundances are in good agreement with the observed values 
in seven BHSXTs.
We emphasize that the abundances vary in a timescale longer than a few months in our model.
Note also that the isotopic ratio \nuc{Li}{6}/\nuc{Li}{7} becomes $0.69 - 0.81$ 
on the secondaries in BHSXTs.
Detection of such a high ratio will be favorable to the production of Li
through the spallation on the secondaries and to the existence of ADAF, 
rather than an accretion disk corona system~\citep[e.g.][]{malzac07}.

Although we have concentrated ourselves on the case of BHSXTs in the present paper,
the scenario for the production of Li is also applicable to the secondaries of NSSXTs,
such as Cen X-4, with a modification on the geometry of ADAF 
due to the magnetic fields of a neutron star.

We can evaluate abundances of Be and B produced via the neutron-induced spallation
and $\gamma$-ray lines emitted through neutron capture on a secondary.
This is our future task.


\end{document}